# Topological Theory of Ceramic High Temperature Superconductors

## J. C. Phillips

Dept. of Physics and Astronomy, Rutgers University, Piscataway, N. J., 08854

## Abstract

**Optimally doped ceramic superconductors (cuprates, pnictides, …) exhibit transition temperatures $T_c$ much larger than strongly coupled metallic superconductors like Pb ($T_c$= 7.2K, $E_g/kT_c$ = 4.5), and exhibit many universal features that appear to contradict the BCS theory of superconductivity based on attractive electron-phonon pairing interactions. Here I argue that this paradoxical simplicity is plausibly resolved within the framework of the Pauling-Phillips self-organized, hard-wired dopant network model of ceramic superconductors, which has previously explained many features of the normal-state transport properties of these materials and successfully predicted strict lowest upper bounds for $T_c$ in the cuprate and pnictide families.**

Ceramic high temperature superconductivity (HTSC) has confounded many theorists (1). Compared to normal metals these "poor metal" ceramics exhibit many transport anomalies even in their normal state, so it is scarcely surprising that their superconductive properties should be even more unexpected. All known microscopic theories of metallic superconductors are based on the BCS model of Cooper pairs formed by attractive electron-phonon interactions that overwhelm repulsive Coulomb interactions, yet many features of electron-phonon interactions in the BCS theory that are confirmed in metals, appear to be weak or absent in the infrared spectra of ceramic HTSC (2). The ceramics often exhibit multiple phases, and must be doped to be not only superconductive, but even metallic. This situation should be compared to the one found in $MgB_2$ with $T_c \sim 40$ K, an "ideal" example of a metallic BCS crystalline superconductor where $T_c$ can be calculated quite accurately, including all the many-body effects associated with strongly coupled electron-phonon interactions (3). When $MgB_2$ is substitutionally doped with $\sim 10\%$ Al (on the Mg sites) or C (on the B sites), $T_c$ decreases drastically to $\sim 10K$, presumably due to inelastic scattering of Bloch waves by dopants (4).

From these contrary chemical doping trends it is already obvious that something drastically different from metallic superconductivity must be happening to cause ceramic HTSC. The most popular explanation for the ceramics is strangely dopant-independent: somehow both phonons and spins synergistically contribute interactions that can form Cooper pairs. However, it has been known for a long time that electron-spin interactions (for instance, with magnetic impurities) drastically suppress $T_c$ by breaking $S = 0$ Cooper pairs (5), and more generally that the insulating antiferromagnetic (AF) and metallic Cooper pair four-particle plaquette channels are completely independent (6). Thus explanations of ceramic HTSC based on "combined" phonon and spin boson exchange (7) require an entirely new microscopic theory of superconductivity, which has so far not appeared. Because ceramics often exhibit multiple phases, much evidence has been adduced for both phonon and spin exchange, but recently the data have shifted overwhelmingly in favor of phonon exchange as the only attractive interaction correlated with $T_c$ and responsible for ceramic HTSC (8).

There is a simple alternative to abandoning BCS, which is consistent with Ockham's razor. We retain the BCS model based only on electron-phonon interactions, but abandon translational invariance and the continuum approximation: in other words, we abandon Bloch plane-wave-like basis states for both electrons and phonons, and suppose that currents are carried by discrete molecular wave packets that percolate coherently from dopant to dopant along filamentary interlayer paths, if necessary avoiding entirely insulating or weakly metallic regions associated with AF or strong Jahn-Teller distortions that destroy large parts of Fermi lines. This "zig-zag" interlayer percolative model (ZZIP) was first proposed in 1989 (9), but unlike Bloch models of the "ideal crystal" BCS cases (Al, Pb, $MgB_2$), it cannot be used to calculate either normal-state or superconductive properties from first principles. However, if one is familiar with the general principles associated with self-organized glassy networks (10,11), one can use homology arguments to describe material properties (such as $T_c$) very economically, and from the (by now large) known ceramic HTSC data base make predictions, certainly one of the most demanding tests of any theory. In this way not only was a master function defining strict least upper bounds on $T_c$ established (12), but this master function very successfully predicted $T_c^{max}$ for the complex oxyhalide $Ba_2Ca_2Cu_3O_6F_2$ (apical F) to be 78 K, in excellent agreement with the highest $T_c$ value for BCCOF reported later, 76 K (13). As shown in Fig. 1, $T_c^{max}(<R>)$ also gives (14) a very good account of the largest $T_c$'s in non-cuprate ceramic HTSC, where $<R>$ is the number of chemical bonds/atom. It is clear that the connectivity and stability of the network are marginal for $<R> = 2$, where the largest $T_c$ is reached. Topological network considerations have explained phase diagrams of superionic conductors (15).

With these predictive successes for $T_c$, already the Pauling-Phillips ZZIP connectivity model has succeeded where any or all combinations of spins and phonons in continuum models have failed. Here we address several further aspects of the ZZIP model, as revealed by new experiments.

**Quantum Percolation**

In a percolation model there are two factors that should be considered: the planar anisotropy of the energy gap $E_g(\phi)$, and the probability $P(\phi)$ that the gap will percolate at a given angle $\phi$. Quantum percolation (8) must occur coherently on each percolative filament, so for quantum percolation one should regard $P(\phi)$ as a probability amplitude, with characteristic analytic properties. Currents will be carried on each filament by wave packets composed of Bloch waves. The wave packets will have average energies E and average velocities **v**. In ARPES experiments these wave packets are projected on the momenta **k** of high energy photoemitted electrons, and the percolative supercurrents are separated from insulating pseudogap states (probably associated with Jahn-Teller distortions) by defining Fermi arcs. When this projective method of analysis is used, it is often difficult to separate the pseudogap and the superconductive gap states, except in underdoped samples where the latter is much smaller than the former (16,17). (At optimal doping, as in (7), the pseudogap and the superconductive gap appear to be nearly the same.) However, even after $E_g(\phi)$ has been assigned a d-wave angular dependence $E_g(\phi) = E_g^{max}\cos 2\phi$, to establish a quantum percolation model one must still determine the probability amplitude $P(\phi)$ that the gap will percolate at a given angle $\phi$. The function $P(\phi)$ is not directly accessible to experiment, but it can be regarded heuristically.

Here $P(\phi)$ is assigned a simple functional dependence for all cuprates, the basic idea being that metallic and superconductive wave packets will percolate most effectively if they avoid the $\phi = 0$ nearest neighbor {10} directions where the pseudogap $E_p(\phi)$ and superconductive gap $E_g(\phi)$ are strongest (18). These are the $\phi = \pi/4$ next nearest neighbor {11} directions where electron-phonon interactions are weakest, thus $P(\phi)$ should be complementary to $E_g(\phi)$, for example, $P(\phi) = \sin^n 2\phi$.

The model is tested by studying the detailed electronic structure of Fermi-energy dopant resonances [type B, predicted in Sec. IX of (12)] as revealed by recent large scale scanning tunneling microscope (STM) experiments (19). As expected, the amplitude of the superconductive gap is greatly reduced in the $\phi = 0$ nearest neighbor {10} directions where the pseudogap $E_p(\phi)$ is strongest. However, only a modest reduction is found in the $\phi = \pi/4$ next nearest neighbor {11} directions where electron-phonon interactions are weakest, which is exactly what one expects from quantum percolation of superconductive wave packets (see Fig. 2). Although this experiment does not display either $E_{p,g}(\phi)$ or $P(\phi)$ fully, it does confirm qualitatively their complementary character.

**Percolative Angular Gap Equation**

The gap equation is obtained by averaging over $\phi$ to obtain $E_g = I_2/I_1$, where $I_2 = \int d\phi E_g(\phi)P(\phi)$ and $I_1 = \int d\phi P(\phi)$ with $0 \le \phi \le \pi/4$. We find $E_g = E_g^{max}/(n + 1)I_1$. At present this equation has limited value, as unambiguous values of $E_g$ and $E_g^{max}$ are not widely available. For example (7) has claimed to have a relation between a possible resonance energy $\Omega_r$, derived from fitting midinfrared data and $T_c$, but (20) obtained quite different resonance energies that show no correlations. The magnetic model of (7) assumes that magnetic dipole resonances can dominate the electric dipole infrared absorption bands, which seems unreasonable, whereas (20) uses a conventional polaronic absorption model.

These ceramic midinfrared data merely reinforce the large qualitative difference between layered ceramics ($La_{2-x}Sr_xCuO_4$) and cubic $Ba_{1-x}K_xBiO_3$ known from their phase diagrams. The latter has a parabolic (nearly semicircular) $T_c(x)$, while the latter has a triangular $T_c(x)$, with a maximum $T_c$ at the metal-insulator transition and a normal BCS-type infrared spectrum with $E_g/kT_c = 3.2$ (21). There have been many studies of the metal-insulator transition at x = 0.3 in $Ba_{1-x}K_xBiO_3$ and these have been interpreted in terms of bipolaron formation (22). However, a more appealing interpretation of the cubic $Ba_{1-x}K_xBiO_3$ metal-insulator transition would be the onset of percolative metallic patches, as the "bipolaronic" infrared dopant peak appears to merge with the CDW band edge at x = 0.3, where $T_c$ is largest. In other words, the two lattice deformations merge to form non-filamentary (but still self-organized) continuum superconductive metallic regions in a first order metal-insulator transition.

Because of the self-organized internal structure, dopant configurations can adapt to thermal fluctuations, and the dissipative scattering (23) characteristic of large gap/$T_c$ ratios (Pb, $T_c$= 7.2K, $E_g/kT_c = 4.5$) is absent from $Ba_{1-x}K_xBiO_3$ ($E_g/kT_c = 3.2$, weak coupling). However, in a ceramic HTSC (17) found $E_g^{max}/kT_c = 10.6 \pm 2.9$, or $E_g/kT_c = (10.6 \pm 2.9)/3I_1 = 4.4 \pm 1.3$ if n = 2. The weak-coupling fit with n = 2 can be justified topologically: suppose each time a wave packet leaves an interlayer dopant to percolate in a metallic plane, it adds a sin2$\phi$ factor to P($\phi$). Then to "reset" itself in an equivalent state, it will have to return to another (statistically equivalent) interlayer dopant, serially experiencing a second factor sin2$\phi$. Thus n = 2 in P($\phi$) could be an intrinsic feature of interlayer effects in the ZZIP model. This value of is $E_g/kT_c$ is similar to that of Pb.

It appears that the layered structure of ceramic HTSC has led to an entanglement of dopant and Jahn-Teller distortions that is much more complex than is apparent from the data on cubic $Ba_{1-x}K_xBiO_3$. This entanglement is imaged in the STM study (Fig. 2) of a rare layered ceramic Fermi-energy pinning resonance (19), which is probably an apical oxygen vacancy. This resonance is stable and observable by STM, whereas most Fermi-energy pinning interstitial O dopant resonances are too fragile to be imaged with a meV potential. It may be that the spatial and spectral entanglement of dopant-centered superconductive gaps with pseudogaps in ceramic superconductors often renders them intrinsically inseparable.

**Footprints of ZZIP**

Although the dopant-centered ZZIP network is apparently not directly observable, it has left many large footprints (12). Thus EXAFS studies of lattice distortions (24) have revealed a third sub-$T_c$ phase transition (in addition to the Jahn-Teller transition at T* and the superconductive transition at $T_c$) shown in Fig. 3. This transition at T = $T_{OP}$ can be explained as a glass transition of the ZZIP network (25); it is especially interesting because, unlike the other two transitions, it cannot be modeled with a Landau order parameter. (It could, however, be compared to formation of Cys-Cys disulfide bonds in proteins.)

Probably the most accurate ceramic HTSC phase diagrams are those obtained by studying the planar resistivity $\varrho_{ab}(T)$, which is most linear in T at and near optimal doping (26,12). The dopant network-forming interactions, although individually weak, are cumulatively very strong because the dielectric energy gained by screening internal electric fields increases as the network conductivity increases due to dopant self-organization (12). The maximization of the T-linearity of $\varrho_{ab}(T)$ near optimal doping (26,12) was the first normal-state transport anomaly to be discovered, but in the intervening 20 years theory has been unable to derive this maximization, and it seems unlikely that this extremal will ever be derived using conventional polynomial methods (for instance, Landau Fermi liquid theory predicts a $T^2$ dependence). This situation is typical of exponentially complex problems which are characteristic of networks, as has been discussed elsewhere (8), but it can be repaired by carefully designed experiments.

In ceramic HTSC self-organization effects manifest themselves by a sharpening of properties characteristic of networks. An obvious and trivial example of such sharpening is the narrowing of the superconductive transition itself, which can be promoted by annealing; this is a small effect for traditional metallic superconductors, but a large effect in ceramic HTSC. If the maximization of the T-linearity of $\varrho_{ab}(T)$ near optimal doping is indeed caused by network self-organization, then that should be manifested by an increase in T-linearity upon annealing. Studies of redistribution of oxygen at a fixed oxygen content in the chain layers of slightly underdoped YBCO with a non-linear $\varrho_{ab}(T)$ showed (27) that annealing for 2-8 hrs at 400-420K does lead to an increase of $T_c$ by about 1K and a reduction of the non-linear component in $\varrho_{ab}(T)$ by ~ 10-15% just above $T_c$.

The classic zero-field work of (26) has recently been extended to high magnetic fields with spectacular results (28). It is found (see Fig. 4) that the $\varrho_{ab}(T)$ regime which is most linear in T occurs in $La_{2-x}Sr_xCuO_4$ at $x = x_c = 0.185(5)$ [shifted from $x = x_c = 0.15$ in zero field], and that this linear regime extends at fixed x vertically all the way from T = 0 right up to the highest T's studied (200K). This T (200K, 0K)-re-entrant result is inexplicable if one thinks of (x,T) = ($x_c$,0) as a "quantum critical point", but it is perfectly understandable in the context of a hard-wired ZZIP network, formed at the annealing temperature (12). In fact, one can go further: the vertical linear T regime is shaped like an hour glass, wide at T = 0 and T = 200K, and narrow at T ~ 100K. This narrowing is due to the formation of pseudogap islands, which first percolate at T* ~ 100K and then stabilize between T* and $T_c$ (see Fig. 3). The low temperature linearity requires a reformation of the nanodomain network into a doubly percolative network, with two kinds of dopants (12).

**Conclusions**

We have seen that the self-organized, hard-wired ZZIP network, although it has eluded direct observation, is still evident in many experiments, where it explains many otherwise inexplicable results. There is an interesting aspect of the hard-wired ceramic HTSC which deserves further thought, although it lies outside the scope of this paper. In spite of considerable interest in quantum computers, their realization has presented many practical problems, especially with regard to programming. Some aspects of this problem might be soluble using the properties of ceramic HTSC, by programming them classically by annealing at high temperatures (say T > 300K) in applied electric and/or magnetic fields, and then operating them at $T < T_c$. One could envision a composite of ceramic HTSC thin films decorated with an array of semiconductor quantum dots, and utilize electron spins in those dots optically to switch phases in the hard-wired ceramic HTSC network (29).

# References

1.  Anderson PW, Lee PA, Randeria M, Rice TM, Trevedi N, Zhang FC (2004) The physics behind high-temperature superconducting cuprates: the 'plain vanilla' version of RVB. *J. Phys.-Cond. Mat.* 16 R755-R769.

2.  Basov DN, Timusk T (2005) Electrodynamics of high-T-c superconductors. *Rev. Mod. Phys.* **77** 721-729.

3.  Liu AY, Mazin II, Kortus J (2001) Beyond Eliashberg superconductivity in $MgB_2$: anharmonicity, two-phonon scattering, and multiple gaps. *Phys. Rev. Lett.* 87, 087005.

4.  Choi HJ, Louie SG, Cohen ML (2009) Anisotropic Eliashberg theory for superconductivity in compressed and doped $MgB_2$. *Phys. Rev. B* 79, 094518 .

5.  Phillips JC *Physics of High-$T_c$ Superconductors* (Academic Press, Boston, 1989).

6.  Dzyaloshinskii IE, Larkin AI (1972) Possible states of quasi-unidimensional systems. *Sov. Phys. JETP* 34 422-427.

7.  Yang J, Hwang J, Schachinger E, Carbotte JP, Lobo RPS, Colson MD, Forget A, Timusk T Exchange boson dynamics in cuprates: optical conductivity of $HgBa_2CuO_{4+\delta}$. *Phys. Rev. Lett.* 102 027003 (2009).

8.  Phillips JC (2009) Prediction of high transition temperatures in ceramic superconductors. *arXiv* 09031306.

9.  Phillips JC (1989) Quantum percolation and lattice instabilities in high-T cuprate superconductors. *Phys. Rev. B* 40 8774-8779.

10. Boolchand P, Lucovsky G, Phillips JC, Thorpe MF (2005) Self-organization and the physics of glassy networks. *Phil. Mag.* 85 3823-3838.

11. Pan Y, Inam F, Zhang M, Drabold DA (2008) Atomistic origin of Urbach tails in amorphous silicon. *Phys. Rev. Lett.* 100 206403.

12. Phillips JC (2007) Self-organized networks and lattice effects in high-temperature superconductors. *Phys. Rev. B* 75 214503.

13. Shimizu S, Mukuda H, Kitaoka Y, *et al*. (2008) Self-doped superconductivity in tri-layered Ba2Ca2Cu3O6F2: A Cu-63-NMR study. *Phys. B* 403 1041-1043.

14. Phillips JC (2009) High temperature cuprate-like superconductivity. *Chem. Phys. Lett.* 473, 274-278.

15. Novita DI, Boolchand P, Malki M, Micoulaut M (2009) Elastic flexibility, fast-ion conduction, boson and floppy modes in AgPO3-AgI glasses. *J. Phys. Cond. Mat.* 21 205106.

16. McElroy K, Lee J, Slezak JA, Lee D-H, Eisaki H, Uchida S, Davis JC (2005) Atomic-scale sources and mechanism of nanoscale electronic disorder in $Bi_2Sr_2CaCu_2O_{8+\delta}$. *Science* 309 1048-1052.

17. Boyer MC, Wise WD, Chatterjee K, *et al*., (2007) Imaging the two gaps of the high-temperature superconductor $Bi_2Sr_2CuO_{6+x}$. *Nature Physics* 3 802-806.

18. Alexandrov AS (2008) Unconventional pairing symmetry of layered superconductors caused by acoustic phonons. *Phys. Rev. B* 77 094502.

19. Chatterjee K, Boyer MC, Wise WD, *et al.* (2008) Visualization of the interplay between high-temperature superconductivity, the pseudogap and impurity resonances. *Nature Physics* 4 108-111.

20. Cojocaru S, Citro R, Marinaro M (2007) Incoherent midinfrared charge excitation and the high-energy anomaly in the photoemission spectra of cuprates. *Phys. Rev. B* 75 220502.

21. Jung CU, Kong JH, Park BH, *et al*., (1999) Far-infrared transmission studies on a superconducting BaPb1-xBixO3 thin film: Effects of a carrier scattering rate. *Phys. Rev. B* 59 8869-8874.

22. Nishio T, Ahmad J, Uwe H (2005) Spectroscopic observation of bipolaronic point defects in $Ba_{1-x}K_xBiO_3$. *Phys. Rev. Lett.* 95 176403.

23. Wada Y (1964) The Effect of Quasiparticle Damping on the Ratio between the Energy Gap and the Transition Temperature of Lead. *Rev. Mod. Phys.* 36 253-257.

24. Zhang CJ, Oyanagi H (2009) Local lattice instability and superconductivity in $La_{1.85}Sr_{0.15}Cu_{1-x}M_xO_4$ (M = Mn, Ni, and Co). *Phys, Rev. B* 79 064521.

25. Phillips JC (2009) Universal Non-Landau, Self-Organized, Lattice Disordering Percolative Dopant Network Sub-Tc Phase Transitions in Ceramic Superconductors *arXiv* 09034376.

26. Ando Y, Komiya S, Segawa K, *et al*. *(*2004) Electronic phase diagram of high-T-c cuprate superconductors from a mapping of the in-plane resistivity curvature. *Phys. Rev. Lett.* 93 267001.

27. Jung J, Abdelhadi MM (2003) Electrical transport and oxygen disorder in YBCO. *Int. J. Mod. Phys.* B 17 3465-3469.

28. Cooper RA, Wang Y, Vignolle B, *et al.* (2009) Anomalous Criticality in the Electrical Resistivity of La$_{2-x}$Sr$_x$CuO$_4$. *Science* 323 603-607.

29. Kroutvar M, Ducommun Y, Heiss D, *et al.* (2004) Optically programmable electron spin memory using semiconductor quantum dots. *Nature* 432 81-84.

This paper is a sequel to Phys. Rev. B 75 214503. The predictions for Tc made there have proved to be accurate far beyond my most optimistic expectations. However, given the novelty of the topological method used, and the controversial character of this subject, I am not optimistic that this paper, which contains a gap equation that yields results in excellent agreement with the most trustworthy experimental value, will be accepted by PRB. Therefore it is formatted for PNAS, where it will be published if it is not accepted. Should it be accepted by PRB, I will of course reformat it.

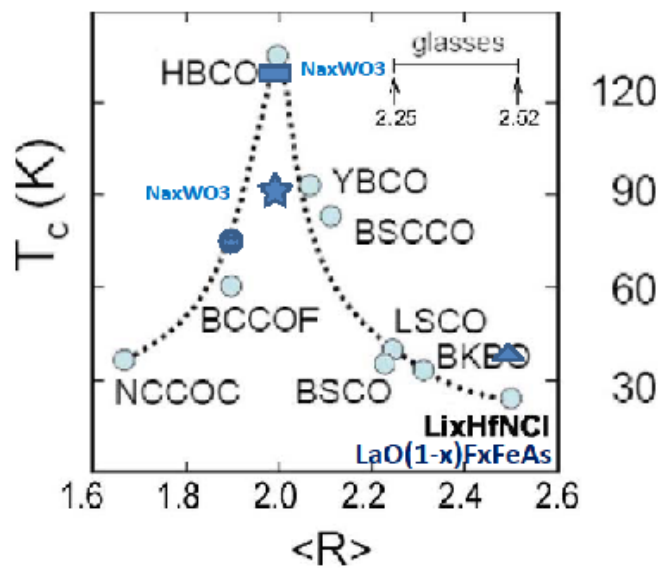

Fig. 1. Master curve for ceramic HTSC, with $T_c^{max}$ plotted as a function of <R> (valence averaged over all atoms) (12). There are two points for the oxyhalide BCCOF, a smaller early value (measured before the master curve was published), and a larger value (measured after the master curve was published). The larger value fits the predicted value almost exactly. Excellent agreement with the predicted values is also found for many other non-cuprate ceramics, including the currently fashionable FeAs superfamily (14).

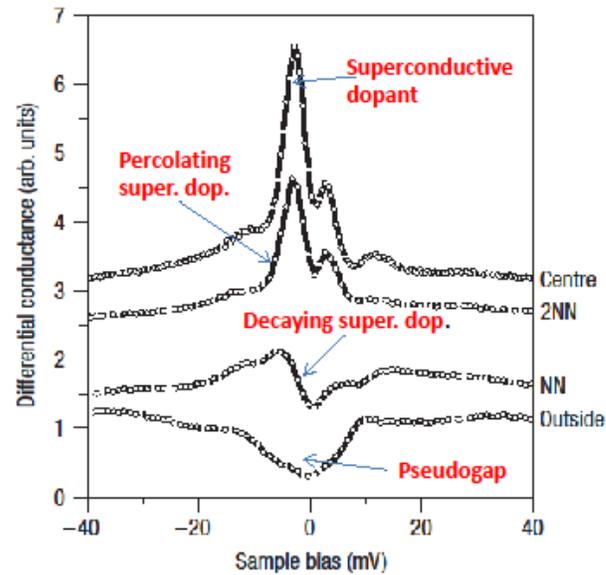

Fig. 2. Images of a Cu site dopant-centered superconductive wave packet (from (19)), qualitatively showing anisotropic percolation favoring {11} second nearest neighbor over decaying nearest neighbor {10} sites, as described quantitatively by the probability amplitude P(φ) in the text. Notice that most of the sample sites far from the dopant are insulating. This means that to thread these regions, interlayer percolation is necessary.

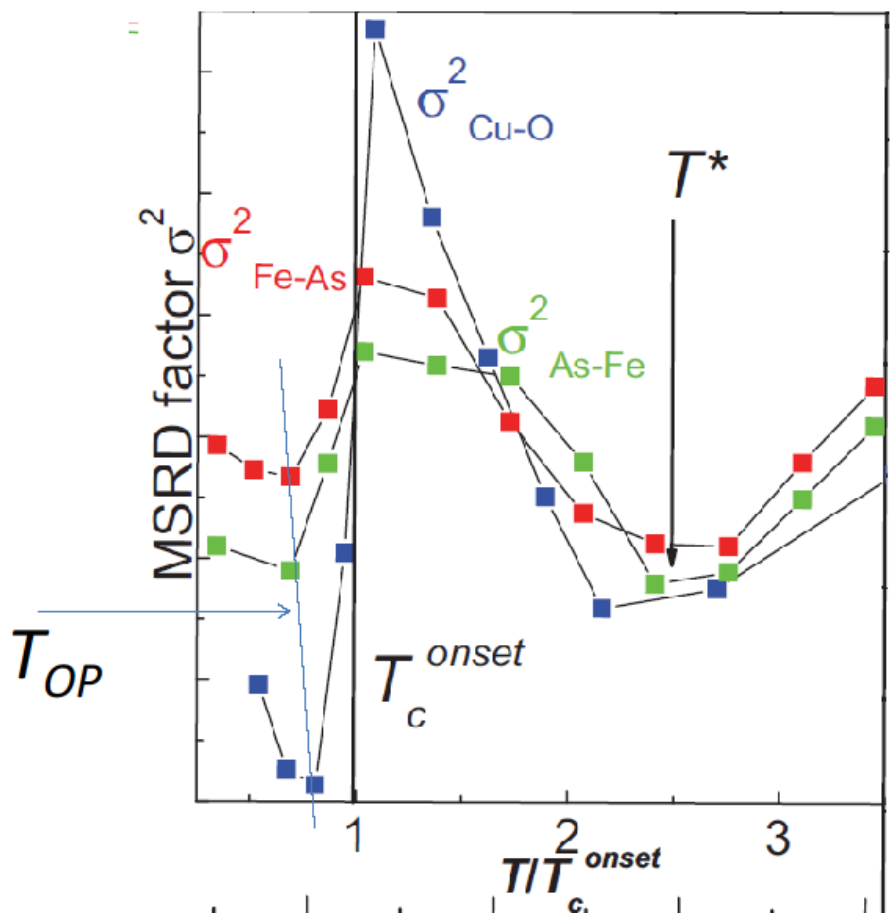

Fig. 3.  EXAFS data (23) showing a sub-$T_c$ phase transition in planar lattice disorder at T = $T_{OP}$. The transition is interpreted in (24) as lattice recoil induced by network superconductive filamentary contacts. Here $T_c^{onset}$ =38K and T* ~ 100K.

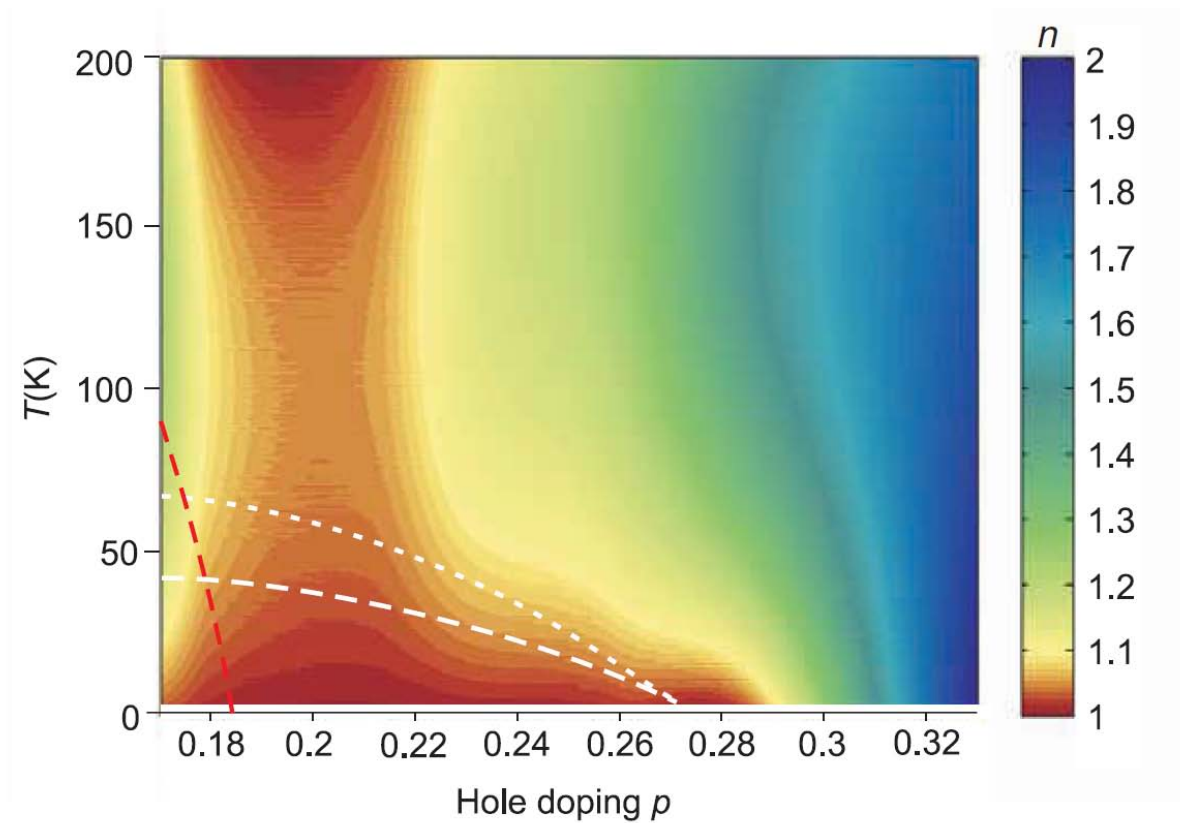

Fig. 4. The exponent n in $\varrho_{ab}(T) \sim T^n$ in $La_{2-x}Sr_xCuO_4$ at high magnetic fields (normal state), from (26). The hour-glass shape of the quasi-linear region n ~ 1.0 (in red) cannot be explained by a quantum critical point at T = 0 or by separation of the Fermi line in momentum space into two components (27), but it is explained by the self-organized formation of the ZZIP network, which is partially disrupted by a Jahn-Teller deformation near T* ~ 100K (see Fig. 3). Notice that at the largest hole doping p > 0.32 the Landau limit n = 2 is reached.